\documentclass[vecphys]{svmult}

\usepackage{makeidx}         
\usepackage{graphicx}        
\usepackage{multicol}        
\usepackage[bottom]{footmisc}
\pdfoutput=1

\makeindex             

\begin{document}

\title*{Spitzer/IRS Mapping of Local Luminous Infrared Galaxies}
\author{M. Pereira-Santaella\inst{1}\and
A. Alonso-Herrero\inst{1}\and
G. H. Rieke\inst{2}\and
L. Colina\inst{1}}
\institute{Instituto de Estructura de la Materia, CSIC, 28006 Madrid, Spain
\and Steward Observatory, University of Arizona, Tucson AZ85721, USA}
%
%
\maketitle
\index{Author1}
\index{Author2}

\begin{abstract}
  
We present results of our program Spitzer/IRS Mapping of local
Luminous Infrared Galaxies (LIRGs). 
The maps cover the central
$20''\times 20''$ or $30''\times 30''$ 
regions of the galaxies, and use all four IRS
modules to cover the full $\sim 5-38\,\mu$m spectral range. We have built
spectral maps of the main mid-IR emission lines, continuum and PAH
features, and extracted 1D spectra for regions of interest in each
galaxy. The final goal is to fully characterize the mid-IR properties
of local LIRGs as a first step to understanding their more distant
counterparts. 
\end{abstract}

\section{Introduction}
\label{sec:0}
Luminous Infrared Galaxies (LIRGs, $L_{\rm IR}=10^{11}-10^{12}L_{\odot}$)
are an important cosmological class of galaxies as they are the main
contributors to the co-moving star formation rate density of the
universe at $z=1$. Moreover, the mid-IR spectra of high redshift
($z\sim 2$) very luminous IR galaxies ($L_{\rm IR}>10^{12}L_{\odot}$)
appear to be better reproduced with those of local starbursts and
LIRGs. This may just reflect the fact that at high-z star-formation
was taking place over a few kiloparsec scales rather than in very
compact ($<1\,$kpc) regions.

We used the spectral mapping capability of
IRS on Spitzer with low (SL+LL modules, $R=60-120$, $5-38\,\mu$m) 
and high (SH+LH
modules, $R=600$, $10-37\,\mu$m) spectral resolution (see \cite{AAH08} for
details) to observe 12 local (distances of 
$<$75Mpc) LIRGs selected from the sample of \cite{AAH06}. 
The maps cover the central
$20''\times 20''$ or $30''\times30''$ regions of the galaxies. 
The spectral data cubes were assembled with CUBISM
\cite{Smith}.


\section{Results}
\subsection{Maps of emission lines and velocity fields}
\label{sec:2}
\begin{figure}
\centering
\includegraphics[width=10.5cm]{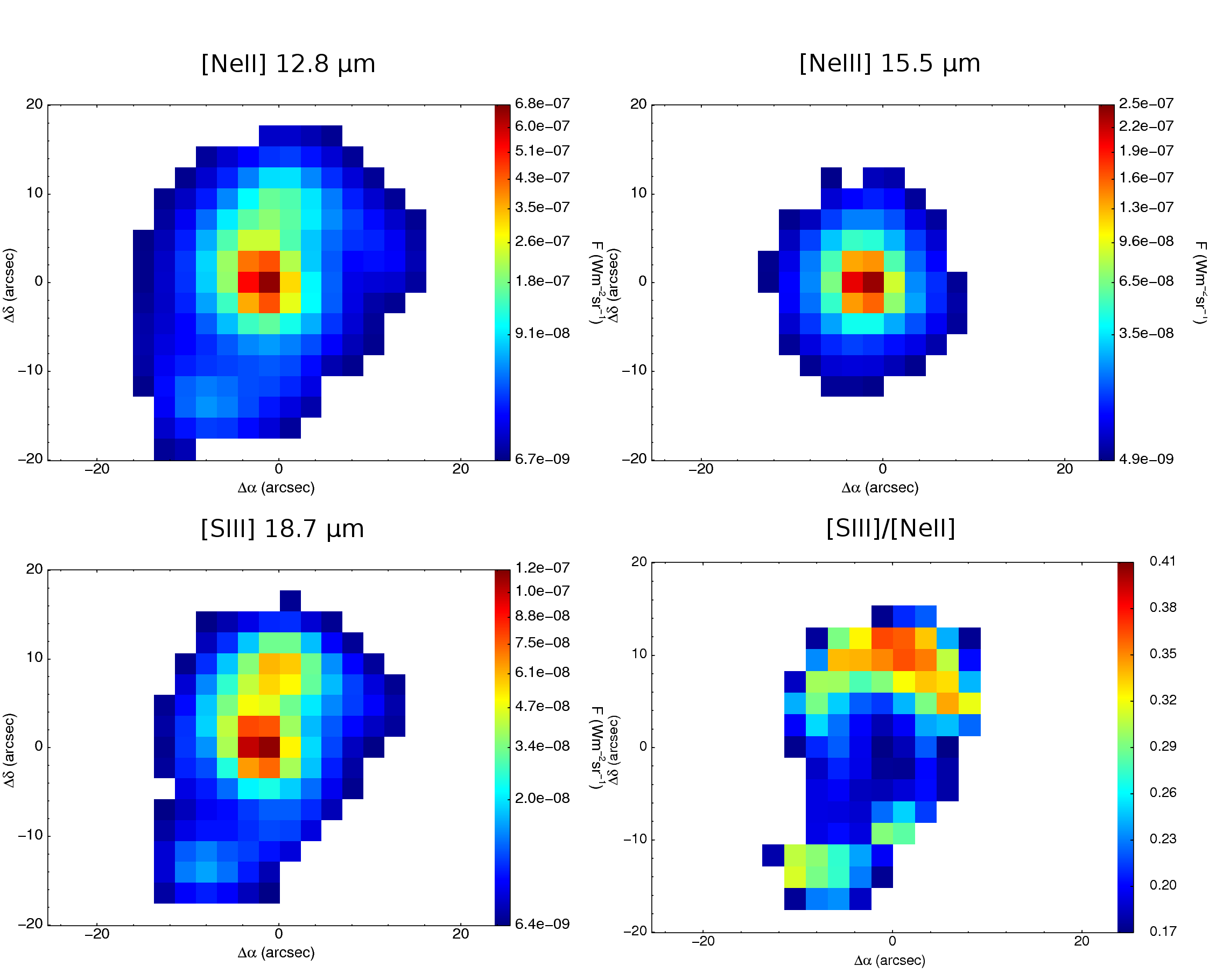}
\caption{Spectral maps and line ratios observed with the SH module 
for NGC~7130.}
\label{fig:1}       
\end{figure}

\begin{figure}
\centering
\includegraphics[width=9.7cm]{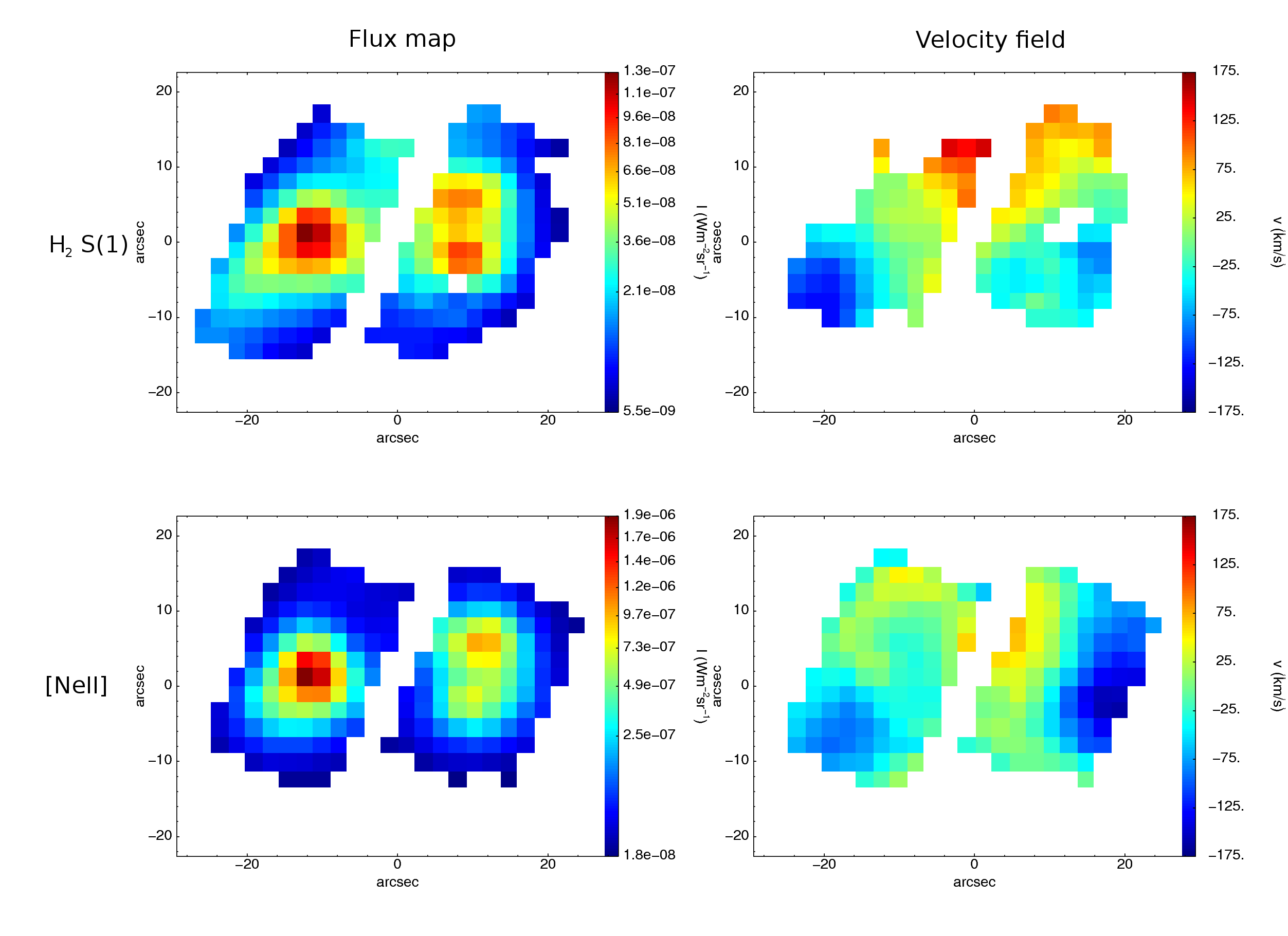}
\caption{Comparison of the SH spectral maps of the flux and velocity
  field of the 17$\mu$m H$_2$ line and the [NeII]12.8$\mu$m line of
  Arp~299.}
\label{fig:2}       
\end{figure}

\begin{figure}
\centering
\includegraphics[width=10.3cm]{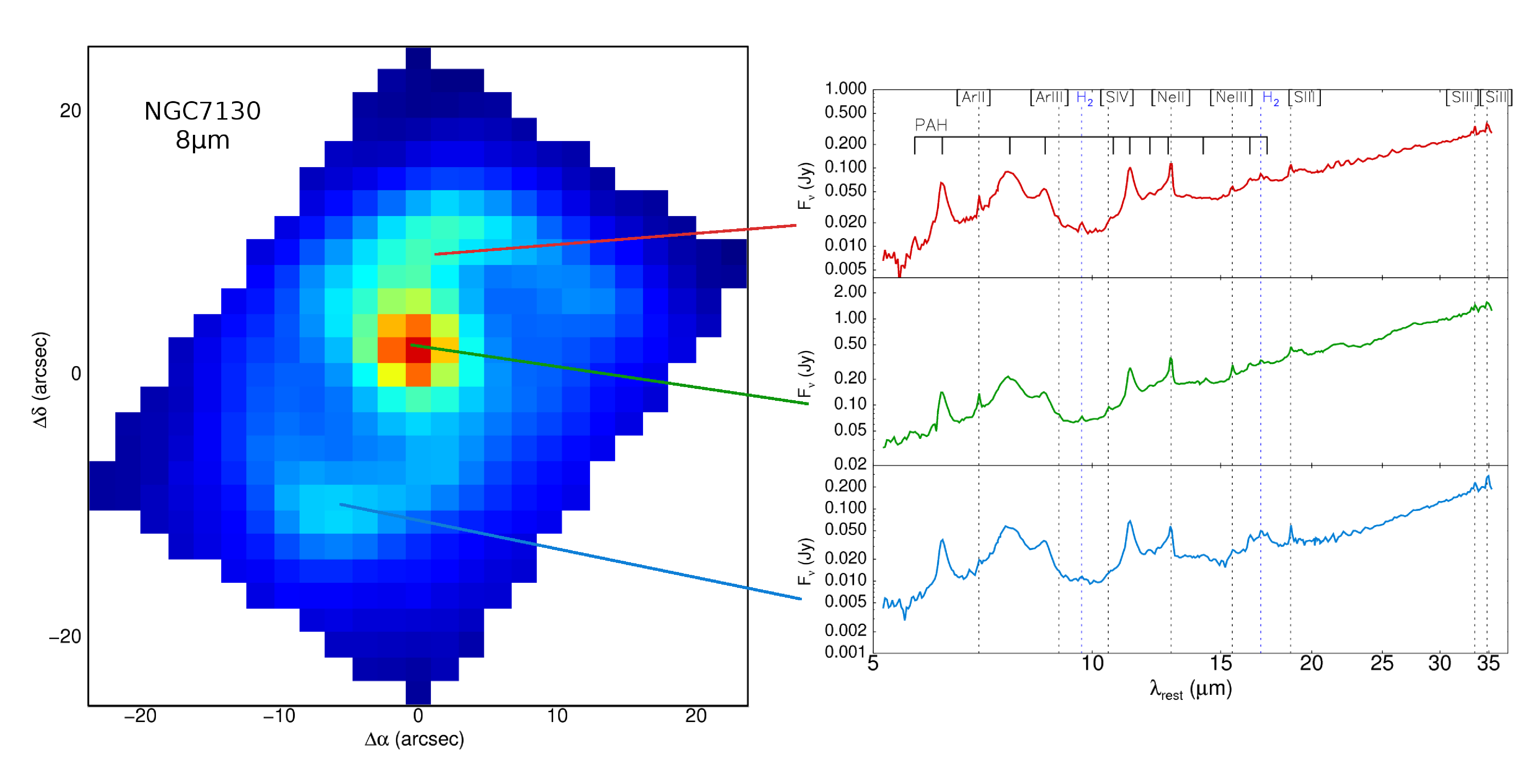}
\caption{{\it Left: }
Spectral map at $8\,\mu$m of NGC~7130 observed with the SL module. 
{\it Right:} 1D spectra observed with the SL and LL modules, 
extracted  for three different regions. The spectra show PAHs, 
emission lines, and the $9.7\,\mu$m silicate feature. }
\label{fig:3}       
\end{figure}

We fitted the brightest emission lines in the 
cubes using Gaussians for the emission lines
and  an order one polynomial for the local continuum. We obtained
spectral maps of the line fluxes, line ratios, FWMHs and velocities (see
Fig.~\ref{fig:1} and Fig.~\ref{fig:2}). 
The spectral maps allow us to select regions of interest in each
galaxy and extract 1D spectra 
to study their physical properties (Fig.~\ref{fig:3}).

\subsection{Silicate absorption}
\label{sec:3}
\begin{figure}
\centering
\includegraphics[width=11.8cm]{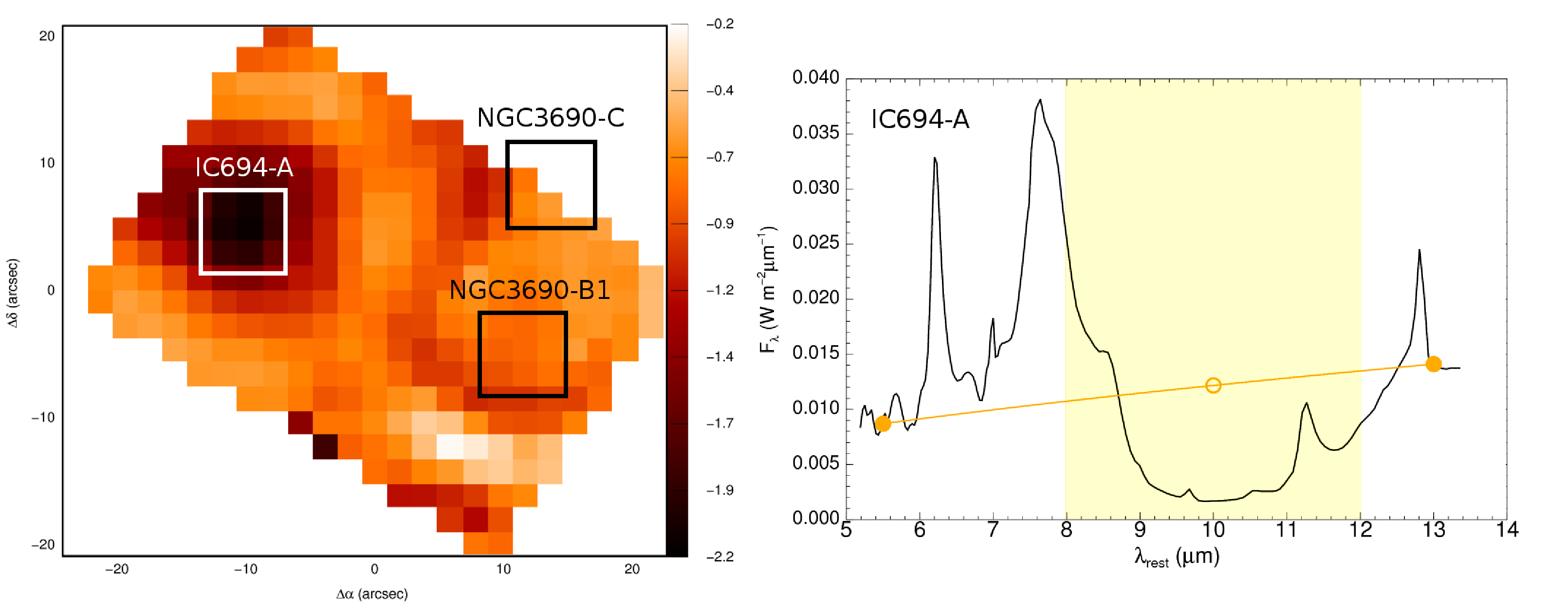}
\caption{\textit{Left}: Map of the strength of the silicate feature 
for the interacting galaxy Arp~299=IC~694+ NGC~3690. \textit{Right}:
Spectrum of the most extincted region, the nuclear region of IC~694,
source A, for which we measured $A_V\sim 34$mag (see \cite{AAH08}).}
\label{fig:4}       
\end{figure}

Amorphous silicate grains have a broad absorption feature centered at
$9.7\,\mu$m. To measure its depth we used a method similar to that
proposed by Spoon et al. \cite{Spoon}. The continuum at $10\,\mu$m is
estimated by fitting a power law through free feature continuum pivots
at $5.5\,\mu$m and $13.0\,\mu$m. The silicate strength is defined as: 
$S_{sil} = ln \frac{f_{obs}(10{\mu}m)}{f_{cont}(10{\mu}m)}$. Assuming
an extinction law and a dust geometry the silicate strength can be
converted into visual extinction. 
Fig.\ref{fig:4} shows the spatial variation of the 
silicate strength in Arp~299.

\subsection{Detecting AGNs}
\label{sec:4}
\begin{figure}
\centering
\includegraphics[width=11.8cm]{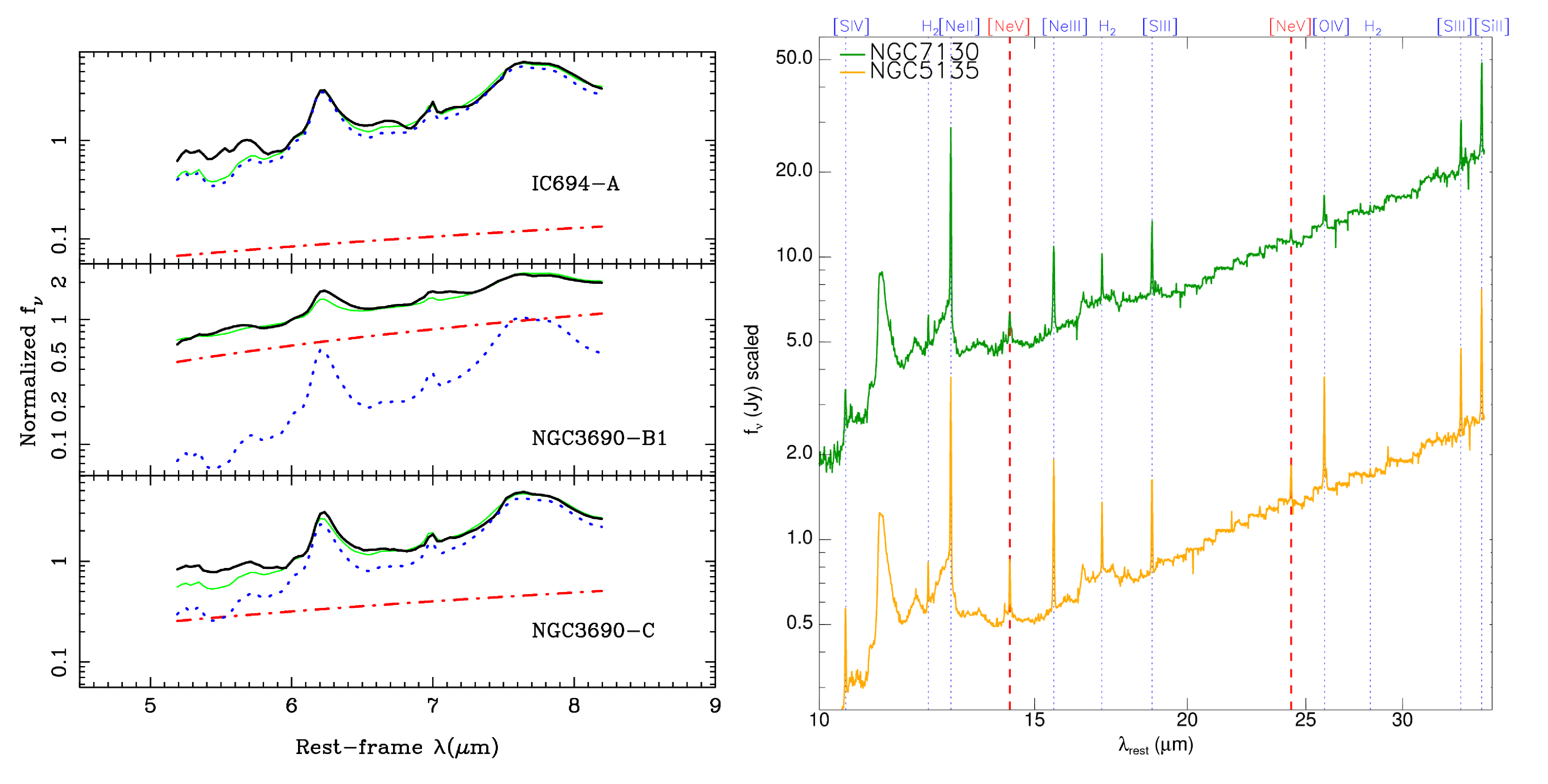}
\caption{\textit{Left}: Spectral decomposition of the three nuclei of
  Arp~299 using the observed SL $5-8\,\mu$m spectra (black lines). 
The dotted blue lines represent the starburst component and the
dot-dash red lines are the AGN component seen as  hot dust continuum. 
The green lines are the sum of the two components. 
An AGN component is clearly detected in the nucleus of NGC~3690, B1 (see
  \cite{AAH08}). \textit{Right}: SH nuclear spectra of NGC~5135 and
  NGC~7130 showing the [NeV] lines as well as the [OIV] line.} 
\label{fig:5}       
\end{figure}

We used two independent methods to detect AGN. The first method is the
detection of high ionization emission lines such as, [NeV] at 14.3 and
24.3$\mu$m (97.1eV) and [OIV] at 25.9$\mu$m (54.9eV). The high
ionization potential of these lines implies the presence of an AGN
because stars cannot generally produce such amount of ionizing
radiation (see Fig.~\ref{fig:5} right). The second method,
proposed by Nardini et al. \cite{Nardini}, uses the $5-8\,\mu$m spectral
region to separate out the starburst component in the form of PAH
emission and the
AGN component in the form of hot dust continuum, as can be seen
from Fig.~\ref{fig:5}. 



\end{document}